\documentclass[iop]{emulateapj}
\usepackage{graphicx}
\usepackage{hyperref}
\usepackage{epsfig}

\def\amin{\ifmmode^{\prime}\else$^{\prime}$\fi}
\def\asec{\ifmmode^{\prime\prime}\else$^{\prime\prime}$\fi}

\def\simgt{\lower.5ex\hbox{$\; \buildrel > \over \sim \;$}}
\def\simlt{\lower.5ex\hbox{$\; \buildrel < \over \sim \;$}}

\def\edot{$\dot E$}

\def\tauc{$\tau_c \equiv P/2\dot P =$ }

\newcommand\chandra{{\it Chandra}}

\newcommand\xmm{{\it XMM-Newton}}

\newcommand\swift{{\it Swift\/}}
\newcommand\nustar{\hbox{\it NuSTAR\/}}
\newcommand\fermi{{\it Fermi\/}}

\slugcomment{Accepted to The Astrophysical Journal}

\begin{document}
\title{A broadband X-ray study of the Geminga pulsar with NuSTAR and XMM-Newton}
\shorttitle{Higher-energy study of Geminga}
\author{Kaya Mori\altaffilmark{1}, Eric V. Gotthelf\altaffilmark{1}, Francois Dufour\altaffilmark{2}, Victoria M. Kaspi\altaffilmark{2}, 
  Jules P. Halpern\altaffilmark{1}, Andrei M. Beloborodov\altaffilmark{1}, Hongjun An\altaffilmark{2}, Matteo Bachetti\altaffilmark{3,4}, 
 Steven E. Boggs\altaffilmark{5}, Finn E. Christensen\altaffilmark{6}, William W. Craig\altaffilmark{5},
 Charles J. Hailey\altaffilmark{1}, Fiona A. Harrison\altaffilmark{7}, Chryssa Kouveliotou\altaffilmark{8},  
Michael J. Pivovaroff\altaffilmark{9}, Daniel Stern\altaffilmark{10}, William W. Zhang\altaffilmark{11}}

\altaffiltext{1}{Columbia Astrophysics Laboratory, Columbia University, New York, NY 10027, USA; kaya@astro.columbia.edu}
\altaffiltext{2}{Department of Physics, McGill University, Montreal, QC H3A2T8, Canada}
\altaffiltext{3}{Universit\'e de Toulouse; UPS-OMP; IRAP; Toulouse, France}
\altaffiltext{4}{CNRS; Institut de Recherche en Astrophysique et Plan\'etologie; 9 Av. colonel Roche, BP 44346, F-31028 Toulouse cedex 4, France}
\altaffiltext{5}{Space Sciences Laboratory, University of California, Berkeley, CA 94720, USA}
\altaffiltext{6}{DTU Space - National Space Institute, Technical University of Denmark, Elektrovej 327, 2800 Lyngby, Denmark}
\altaffiltext{7}{Cahill Center for Astronomy and Astrophysics, California Institute of Technology, Pasadena, CA 91125, USA}
\altaffiltext{8}{NASA Marshall Space Flight Center, Huntsville, AL 35812, USA}
\altaffiltext{9}{Lawrence Livermore National Laboratory,  Livermore, CA 94550, USA}
\altaffiltext{10}{Jet Propulsion Laboratory, California Institute of Technology, Pasadena, CA 91109, USA}
\altaffiltext{11}{NASA Goddard Space Flight Center, Greenbelt, MD 20771, USA}

\shortauthors{Mori et al}

\begin{abstract}
We report on the first hard X-ray detection 
of the Geminga pulsar above 10 keV using a 150~ks observation with the \nustar\ observatory. The double-peaked pulse profile  
of non-thermal emission seen in the soft X-ray band persists at higher energies.  
Broadband phase-integrated spectra over the 0.2--20 keV band with \nustar\ and archival \xmm\ 
data do not fit to a conventional two-component model of a blackbody plus power-law, but instead exhibit spectral hardening above $\sim5$ keV. 
We find two spectral models fit the data well: (1) a blackbody ($kT_1 \sim 42 $ eV) with a broken power-law
($\Gamma_1 \sim 2.0$, $\Gamma_2 \sim 1.4$ and $E_{\rm break}\sim3.4$ keV), and (2) two blackbody components ($kT_1 \sim 44 $ eV and $kT_2 \sim 195 $ eV) with a power-law 
component ($\Gamma \sim 1.7$). 
In both cases, the extrapolation of the Rayleigh-Jeans tail of the thermal component is consistent with the UV data, while the non-thermal component overpredicts the near-infrared data,  
requiring a spectral flattening at $E\sim0.01-1$ keV. 
While strong phase variation of the power-law index is present below $\sim5$ keV, our phase-resolved spectroscopy with \nustar\ indicates that another  
hard non-thermal component with $\Gamma\sim1.3$ emerges above $\sim 5$ keV.
The spectral hardening in non-thermal X-ray emission as well as spectral flattening between the optical and X-ray bands argue against 
the conjecture that a single power-law may account for multi-wavelength non-thermal spectra of middle-aged pulsars. 
\end{abstract}


\section{Introduction}
\label{intro}
Geminga was discovered as a bright GeV source by the \textit{SAS-2} experiment \citep{thompson_1977}. Later, the \textit{ROSAT} X-ray observatory identified it as a pulsar with a 237~ms spin period and a 
soft thermal spectrum with a blackbody 
temperature $kT\sim40$ eV \citep{halpern_1992, bignami_1992}. The Energetic Gamma-Ray Experiment Telescope (EGRET) confirmed the  
pulsations \citep{bertsch_1992} and measured the pulsar spin-down, establishing that 
Geminga is a rotation-powered pulsar with a spin-down age \tauc $ 3.4\times10^5$ years, 
a spin-down power \edot $ = 3\times10^{34}$ erg~s$^{-1}$ and a dipole 
magnetic field strength $B = 1.6\times10^{12}$~G. 

Over the last two decades, Geminga has been observed and studied in multi-wavelength bands from radio
to TeV (see \citet{bignami_1996} for a review).
The Geminga pulsar stands out among thousands of pulsars because it is the second brightest 
gamma-ray source in our Galaxy with nearly 90\% gamma-ray radiation efficiency ($L_{\gamma}/$\edot) \citep{caraveo_2013}. Its gamma-ray spectrum is 
well described by a power-law with photon index $\Gamma = 1.3$ and an exponential cut-off at 
$E_c = 2.5$ GeV \citep{abdo_2010}. The gamma-ray emission has been mostly 
attributed to curvature radiation from relativistic electrons or inverse 
Compton scattering in the outer gap formed near the light cylinder \citep{cheng_1986, romani_1996, harding_2008, lyutikov_2013}.   

After the discovery of pulsations by \textit{ROSAT}, \textit{ASCA} revealed a hard 
non-thermal component with a power-law index $\Gamma\sim1.5$ extending to 10~keV 
\citep{halpern_1997}. 
X-ray spectra of middle-aged rotation-powered pulsars are often composed of thermal and non-thermal emission (e.g., Geminga, PSR B0656+14 and PSR B1055$-$52; \citet{deluca_2005}). 
The bulk of thermal emission from the neutron star (NS) surface is likely due to heat transfer from the NS interior,  
while non-thermal emission comes from synchrotron radiation in the magnetosphere. 
A phase-resolved spectroscopic study using deep \textit{XMM-Newton} observations argued for the 
presence of a second thermal component with a blackbody temperature of $kT\sim190$ eV 
\citep{caraveo_2004} associated with polar caps heated by returning current from the magnetosphere or due to anisotropic heat conduction in the NS crust \citep{greenstein_1983}. However, \citet{jackson_2005} disputed 
this claim because phase variation of the non-thermal component can account for the phase-resolved spectra without requiring a second blackbody 
component. In either case, a second thermal component of the Geminga pulsar, if it exists, is nearly 
two orders of magnitude fainter than those of two other middle-aged pulsars, PSR B0656+14 
and PSR B1055$-$52 \citep{deluca_2005}.     

Geminga has also been detected at near-infrared (NIR) to UV wavelengths 
\citep{bignami_1993, caraveo_1996, shibanov_2006, danilenko_2011}, exhibiting two components - a power-law spectrum with 
$\Gamma \sim 1.4$ and the Rayleigh-Jeans (RJ) tail of the thermal emission detected in the X-ray band \citep{kargaltsev_2005}. \citet{pavlov_1996} demonstrated that joint UV and X-ray spectroscopy is a powerful 
diagnostic tool to constrain the NS atmospheric composition. 
 
Despite a long-term multi-wavelength observation campaign, the hard X-ray emission (10-100 keV) from Geminga remained 
undetected due to the lack of sensitive hard X-ray telescopes.  
In this paper we report on hard X-ray observations of the Geminga pulsar by the \textit{Nuclear Spectroscopic Telescope Array} (\nustar; \citet{harrison_2013}). 
\nustar\ provides the most sensitive probe to date of the Geminga pulsar above 10 keV, with negligible 
contamination from its faint pulsar wind nebula (PWN) discovered by \chandra\ and \xmm\ \citep{caraveo_2003, pavlov_2006, deluca_2006, pavlov_2010}. 
With broadband spectroscopy from \nustar, archival \xmm\ data and the published results in NIR to UV bands,    
we report on new constraints on both the thermal and the non-thermal emission. We use the parallax distance of 250$^{+120}_{-62}$ pc  
\citep{faherty_2007}, updated from \citet{caraveo_1996}, to rule out several thermal models and calculate X-ray luminosities. 

The paper is organized as follows. \S \ref{nu_obs} and \ref{XMM_obs} present the set of observations 
used and describes our data reduction for \nustar\ and \xmm, respectively.  
In this paper, we analyzed 15 \nustar\ observations and 9 archival \textit{XMM-Newton} observations (Table \ref{observations_table}). 
\S \ref{spectro} presents phase-integrated spectroscopy using \nustar\ and 
\xmm\ data jointly. \S \ref{phase_spec} presents the \nustar\ detection of the pulsations above 10 keV. We compare \nustar\ pulse profiles with those from \xmm\ and \fermi\ and study phase variation of the 
thermal and non-thermal components.  
\S \ref{discussion} summarizes our results and discusses their implications for the thermal 
and non-thermal emission mechanisms of the Geminga pulsar and rotation powered pulsars in general. In the Appendix, we show that the absolute \nustar\ timestamp is accurate to better than 3 ms, and present a new ephemeris 
of Geminga based on \xmm\ and \fermi\ data. 


\begin{deluxetable*}{cccc}
 \tablewidth{5in}
 \tablecaption{Observation log of Geminga pulsar}
 \tablehead{\colhead{Obs.ID} & \colhead{Start Date} & \colhead{Instrument} & \colhead{Net Exposure (ks)}}
 \startdata
 \multicolumn{4}{c}{\textit{NuSTAR}} \\ \hline
 30001029002 & 2012-09-20 & FPMA/FPMB & 7.26 \\
 30001029004 & 2012-09-20 & FPMA/FPMB & 8.81 \\
 30001029006 & 2012-09-20 & FPMA/FPMB & 9.43 \\
 30001029008 & 2012-09-21 & FPMA/FPMB & 4.82 \\
 30001029010 & 2012-09-21 & FPMA/FPMB  & 4.95 \\
 30001029012 & 2012-09-21 & FPMA/FPMB & 13.8 \\
 30001029014 & 2012-09-25 & FPMA/FPMB & 2.43 \\
 30001029016 & 2012-09-25 & FPMA/FPMB & 4.36 \\
 30001029018 & 2012-09-25 & FPMA/FPMB & 26.5 \\
 30001029020 & 2012-09-26 & FPMA/FPMB & 6.52 \\
 30001029022 & 2012-09-26 & FPMA/FPMB & 20.9 \\
 30001029024 & 2012-09-27 & FPMA/FPMB & 5.18 \\
 30001029026 & 2012-09-27 & FPMA/FPMB & 4.33 \\
 30001029028 & 2012-09-27 & FPMA/FPMB & 23.3 \\
 30001029030 & 2012-09-28 & FPMA/FPMB & 5.00 \\ \hline
 \multicolumn{4}{c}{\textit{XMM-Newton}\tablenotemark{a}} \\ \hline
 0111170101 & 2002-04-04 & PN/MOS1/MOS2 & 58.5/80.4/80.1 \\
 0201350101 & 2004-03-13 & PN/MOS1/MOS2 & 12.3/16.5/17.1 \\
 0301230101 & 2005-09-16 & PN\tablenotemark{b} & 6.59 \\
 0311591001 & 2006-03-17 & PN/MOS1/MOS2 & 19.0/25.2/23.8 \\
 0400260201 & 2006-10-02 & PN/MOS1/MOS2 & 13.7/18.9/18.9 \\
 0400260301 & 2007-03-11 & PN/MOS1/MOS2 & 15.0/20.5/20.0 \\
 0501270301 & 2008-03-08 & PN/MOS1/MOS2 & 7.98/10.9/10.6 \\
 0550410201 & 2008-10-03 & PN/MOS1/MOS2 & 14.0/18.6/17.6 \\
 0550410301 & 2009-03-10 & PN/MOS1/MOS2 & 7.40/10.0/9.23
 \enddata
 \tablenotetext{a}{In all \xmm\ observations, the PN camera was operated in Small Window mode with the thin filter, and 
both MOS cameras were operated in Full Frame  mode with the medium filters. We did not use \xmm\ data from observation 0501270201 in 
September 2007 due to an attitude reconstruction issue.}
 \tablenotetext{b}{MOS cameras were not operating in this observation.}
 \label{observations_table}
\end{deluxetable*}


\section{\nustar\ observations}
\label{nu_obs}

The \nustar\ telescope consists of two
coaligned telescopes with corresponding focal plane modules A and B (FPMA and
FPMB).  These modules have an angular resolution of 18\asec~FWHM and 58\asec~half power diameter, and an energy
resolution of 400 eV (FWHM) at 10 keV \citep{harrison_2013}.  The
nominal energy band of \nustar\ is 3--79 keV.  The relative timing
accuracy of the \nustar\ timestamps is \hbox{$\sim2$~ms} after
correcting for thermal drift of the spacecraft clock. In Appendix B 
we show that the absolute Barycentric Dynamical Time (TDB) timestamp
is accurate to better than $\sim3$~ms.

A \nustar\ observing campaign of the Geminga pulsar was carried out
on 2012 Sept 20-28 in a series of 15 short pointings;
an observation log is presented in Table
\ref{observations_table}. 
Data products and response files were generated using NuSTARDAS
v.1.2.0. The filtered event files produce a total of $148$~ks good
exposure time, varying from $2.4$~ks to $26.5$~ks between pointings.
In this work, we limit our analysis from 3 to 20 keV due to low
signal-to-noise of the Geminga data at higher energies. Since the
source is not variable between the short pointings, 
we analyzed images and spectra from the combined data set.  We used
the HEASARC software version 6.13 to analyze both \nustar\ and
\xmm\ data sets.

\subsection{Imaging analysis}

\begin{figure*}[t]
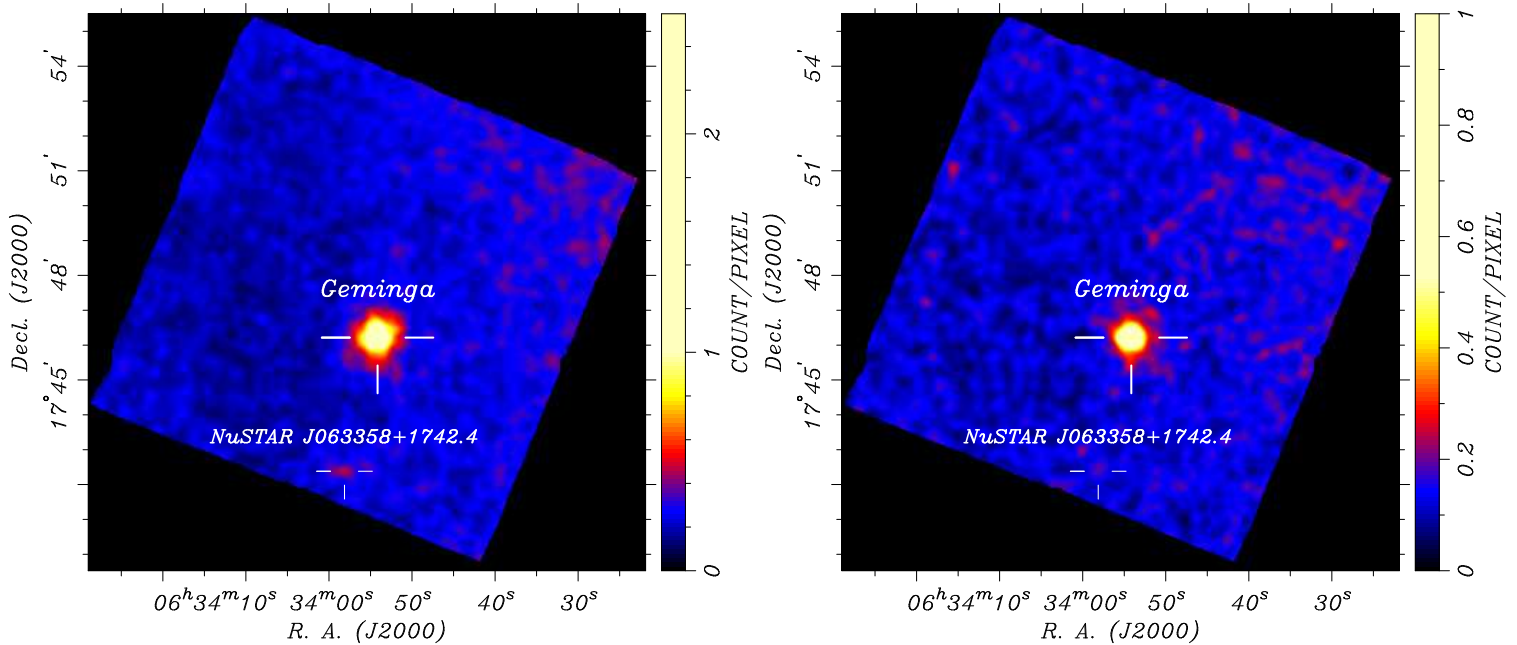

\centerline{
\hfill
\psfig{figure=geminga_3_10keV_trim_color.ps, height=0.55\linewidth, angle=270}
\hfill
\psfig{figure=geminga_10_20keV_trim_color.ps, height=0.55\linewidth, angle=270}
\hfill
}
\caption{\nustar\ mosaic image of the field containing Geminga pulsar, in the 3-10 keV (left) and 10-20 keV (right) bands. 
Data from all Geminga \nustar\ observations and modules are merged, with appropriate exposure
corrections applied.
The locations of Geminga and NuSTAR J063358+1742.4, as
determined using \textbf{wavedetect}, are indicated by the tick marks.}
\label{images}
\end{figure*}

First, we constructed mosaic maps by combining data from all the observations following \citet{nynka_2013}. 
We registered the known {\it HST} position of the Geminga pulsar \citep{faherty_2007}. 
We generated exposure maps using \textbf{nuexpomap} and merged the exposure-corrected images.
Figure \ref{images} shows the mosaic image of the Geminga pulsar in the 3-10 and 10-20 keV energy bands. 
Given \textit{NuSTAR}'s angular resolution, the bright source in Figure \ref{images} is 
consistent with being a point source. The pulsar was not visible above 20 keV, the point at which the instrument background exceeds the source strength. 

We searched for serendipitous point sources in the \nustar\ mosaic images using \textbf{wavdetect}. In the 3-10 keV band, we detected a 
known narrow-line AGN at $z=0.891$ (NuSTAR J$063358+1742.4$) $\sim4'$ south of the Geminga pulsar \citep{alexander_2013}. 
None of the faint PWN features around the pulsar \citep{pavlov_2010} are visible 
because they are swamped by the brighter pulsar emission and \textit{NuSTAR} background photons.

\subsection{Spectral extraction and nebular contamination}

For spectral analysis, we extracted source spectra from a circular aperture with $30\arcsec$ radius centered at the Geminga pulsar position and background spectra using a $60\arcsec < r < 120\arcsec$ annulus.     
Since each detector chip has a different internal background level, it is important to extract background spectra 
from the same detector chip where the source was located (detector 0 in all observations of Geminga).  
To assess the robustness of our results with respect to the choice of background region, 
we also extracted background spectra from several circular regions with radius $90\arcsec$ on detector chip 0, 
finding the results only changed by a small fraction of the statistical uncertainty. 
For the subsequent spectral analysis, we limit the energy band to below 20~keV,
above which the background becomes dominant with several strong emission lines at $\sim$20-30 keV. 

To assess contamination from the PWN, we estimated \textit{NuSTAR} count rates using Table 1 in \citet{pavlov_2010} for the brightest PWN feature, the ``A-tail''. We predict $5\times10^{-4}$ (3-10 keV) and $3\times10^{-4}$ cts~s$^{-1}$ (10-30 keV) within an $r=30\arcsec$ (equal to the Half Power radius) circle.
These estimated count rates are $\sim$5-7 times smaller than the combined count rate of the \textit{NuSTAR} background
and pulsar at the location of the PWN.
Furthermore, using archival \textit{Chandra} data from 2012 and 2013, we found that the overall contamination from the PWN features
within 30 arcsec of the pulsar  is less than 5\% and is thus below our statistical uncertainties 
(see \citet{pavlov_2006, deluca_2006, pavlov_2010} for the individual PWN features and their positions/fluxes).

\section{\xmm\ observations}
\label{XMM_obs}

We analyzed nine archival \xmm\ observations of the Geminga pulsar from
2002 to 2009 (see Table \ref{observations_table}), reduced using the
Standard Analysis Software {\it SAS v.12} and the most up-to-date
calibration files. The EPIC-PN data, acquired in high time-resolution
{\tt SmallWindow} mode (6~ms readout), are most suitable for our timing
and spectral analysis; the 0.3~s EPIC-MOS data suffer from photon
pile-up and is not used here.  The data reduction and analysis follow
the studies of \citet{caraveo_2004}, \citet{jackson_2005} and \citet{deluca_2005}. After filtering out
background flares using the EPIC-PN count rate threshold of $0.05$
cts~s$^{-1}$ above 10 keV, we obtained a total exposure time of
$154.5$~ks. 

For spectral analysis, we extracted source counts from a circular
aperture with a radius of $15\arcsec$ centered on the source position,
computed using the {\it SAS} \textbf{emldetect} routine.  We choose a
small source extraction compared to previous studies to optimize the high energy
($>4$~keV) signal-to-noise ratio at the expense of the low energy
throughput. Background spectra are extracted from a region with a
radius of $30\arcsec$ placed at the same CCD column for each
observation, to avoid the faint PWN features \citep{pavlov_2006}.  We
combined all \xmm\ EPIC-PN spectra using the {\tt FTOOL }\textbf{addascaspec} and performed spectral fitting in the 0.25-10~keV band.


\section{Joint spectral analysis with \nustar\ and \xmm}
\label{spectro}

We analyzed \nustar\ and \xmm\ spectra of the Geminga pulsar by jointly fitting multiple spectra using XSPEC 12.8.  
We grouped each spectrum by a minimum of 30 counts per bin. 
When we fit multiple data jointly, we used a multiplicative factor (\textbf{const} command in XSPEC) for each data set in order to take into account the small flux 
calibration errors ($<10$\%) between \xmm\ and \nustar. We adopted 1-$\sigma$ (68\% c.l.) errors for all the spectral fitting results presented in this paper.    
In order to distinguish between different spectral models, we used $\chi^2$ statistics and the F-test (\textbf{ftest} command in XSPEC) as a null hypothesis test.

\subsection{Non-thermal spectral fitting above 3 keV}
\label{spec_3keV}

\begin{figure}
\centerline{
\hfill
\includegraphics[width=0.7\linewidth, angle=270, clip=]{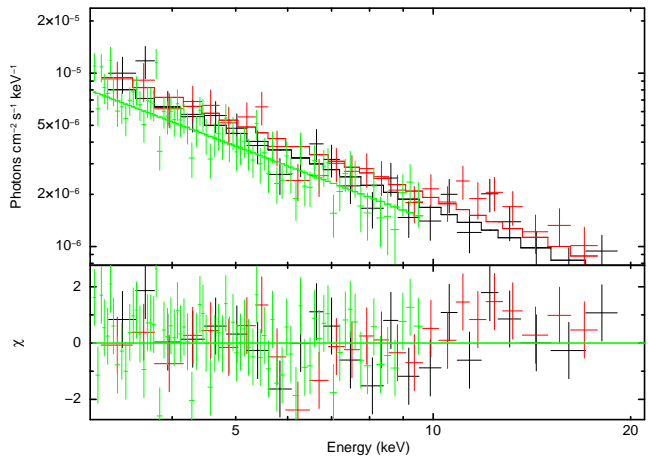}
\hfill
}
\caption{Unfolded 3-20 keV \nustar\ + \xmm\ EPIC-PN spectra of the Geminga pulsar fit with a single power-law model. The green, black and red data points are for \xmm\ EPIC-PN, \nustar\ FPMA and FPMB, respectively.} 
\label{fig:nustar_xmm_spec}
\end{figure}

Thermal and non-thermal components in pulsar X-ray spectra can be strongly covariant with inherent  parameter degeneracy. In order to constrain the non-thermal 
component cleanly, we analyzed the X-ray spectra above 3 keV where the contribution from the second blackbody component is negligible \citep{deluca_2005}.  
We note that the low absorption column ($\sim10^{20}$~cm$^{-2}$) does not affect the Geminga spectra above 3 keV. 
The 3-20 keV \nustar\ spectra are well fit to a single power-law model with $\Gamma = 1.35\pm0.08$ and a reduced $\chi^2$ of 0.97 (49 dof). 
In order to improve the photon statistics, we jointly fit 3-20 keV \nustar\ spectra and 3-10 keV 
\xmm\ EPIC-PN spectra. Figure \ref{fig:nustar_xmm_spec} shows the spectral fits and residuals, 
yielding $\Gamma = 1.44\pm0.06$ and a reduced $\chi^2$ of 0.97 (121 dof). 
We find that a single power-law fit is adequate; any additional continuum component such as a blackbody 
or a second power-law model did not improve the spectral fit significantly, with F-test false probabilities of $\sim0.02$. 
The power-law index from the 3-20 keV spectra is consistent with that of \citet{kargaltsev_2005} ($\Gamma = 1.56\pm0.24$) where they fit \xmm\ EPIC-PN spectra 
in 2.5-10 keV, but our broadband data provide stricter constraints.  

\subsection{Broadband spectral fitting between 0.2 and 20 keV}
\label{nustar_xmm}

\begin{figure}
\centerline{
\hfill
\psfig{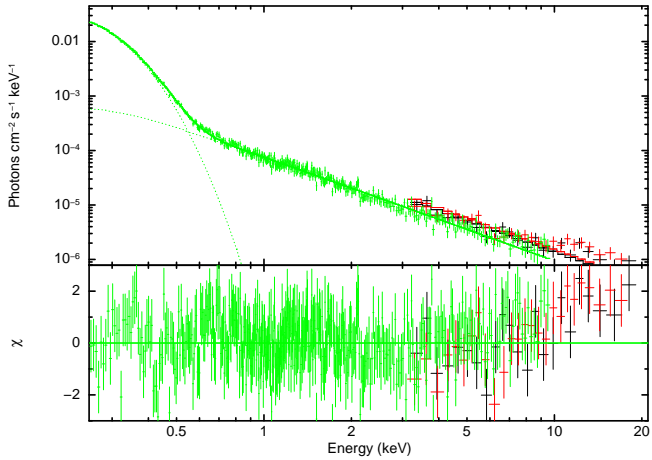}
\hfill
}
\caption{Unfolded 0.25-20 keV \nustar\ and \xmm\ (EPIC-PN) spectra of the Geminga pulsar fitted with the BB+PL model. Residuals are seen especially in the \nustar\ data above 10 keV. }
\label{fig:bb_pl_spec}
\end{figure}

We analyzed broadband X-ray spectra of the Geminga pulsar in the 0.25-20 keV band with 3-20 keV \nustar\ spectra and 0.25-10 keV \xmm\ EPIC-PN spectra. 
We use the \textbf{tbabs} absorption model in XSPEC, with Wilms abundances and Verner cross-sections \citep{wilms, verner}.
First, we fit a blackbody plus power-law (BB+PL) model to the \nustar\ plus \xmm\ EPIC-PN spectra. The fit parameters are consistent with the previous analysis of \citet{jackson_2005} (Table \ref{index_table}). However, 
some residual excesses are clearly seen above $\sim5$ keV, indicating the presence of an additional spectral component (Figure \ref{fig:bb_pl_spec}). 
The fit power-law index ($\Gamma = 1.90\pm0.02$) is softer than that from the 3-20 keV spectral analysis 
($\Gamma = 1.44\pm0.06$). A similar discrepancy in the fit power-law index between the entire band and high energy band was previously reported by \citet{kargaltsev_2005} 
using only \xmm\ EPIC-PN data. 

Instead of a blackbody model, we also fit a magnetized hydrogen NS atmosphere model for the surface magnetic field $B = 10^{12}$ G 
(\textbf{nsa} in XSPEC) \citep{pavlov_1995}. For a given effective temperature, a NS hydrogen atmosphere spectrum is harder than a blackbody because the dominant free-free absorption opacity  decreases 
with photon energy. Although the fit quality and residuals are similar to those of the blackbody model fit, the NS hydrogen atmosphere model yields a radius ($R=440\pm60$~km) that is inconsistent with any proposed 
models for the NS equation of state \citep{lattimer_2007}. In addition, the RJ tail in the UV band is significantly over-estimated because the thermal flux in the RJ tail is proportional to $R^2 \times T$  
\citep{kargaltsev_2005}. Therefore we rule out the NS hydrogen atmosphere model, and hereafter we use only a blackbody as a thermal model.   

\begin{figure*}[t]
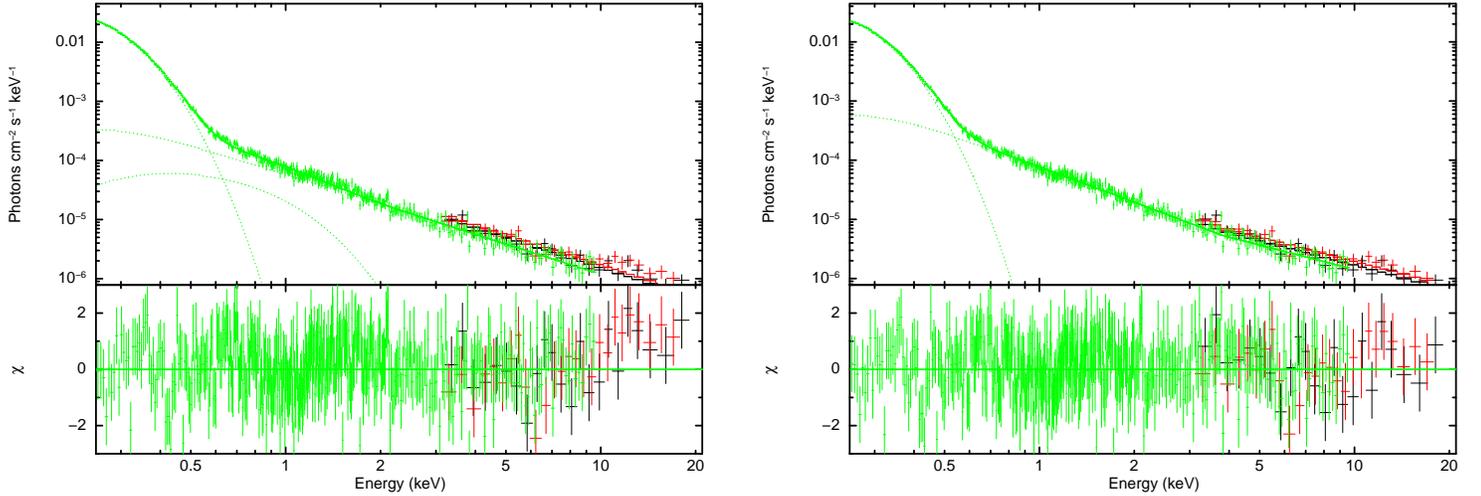

\centerline{
\hfill
\psfig{figure=2bb_pl_unfold.ps,height=0.55\linewidth, angle=270}
\hfill
\psfig{figure=bknpo_bb_unfold.ps,height=0.55\linewidth, angle=270}
\hfill
}
\caption{Unfolded 0.25-20 keV \nustar\ and \xmm\ (EPIC-PN) spectra of the Geminga pulsar fitted with 2BB+PL (left) and BB+BKPL model (right). }
\label{fig:nustar_xmm_spec2}
\end{figure*}

Given the excess residuals from fitting the BB+PL model, we fit three different models either 
adding a blackbody or power-law model or using a broken power-law as the non-thermal component. 
Specifically, we fit two blackbodies plus a power-law (2BB+PL), a blackbody plus two power-laws (BB+2PL) and a blackbody plus broken power-law (BB+BKPL, with \textbf{ bknpower} model 
in XSPEC). All the three models greatly improved the spectral fit compared to the BB+PL model with F-test false probabilities $< 10^{-7}$. Figure \ref{fig:nustar_xmm_spec2} shows the \nustar\ 
and \xmm\ spectra fit by the 2BB+PL and BB+BKPL models. All three models (BB+BKPL, BB+2PL and 2BB+PL) agree reasonably well with the UV data (i.e. the RJ tail of the X-ray thermal emission). 
However, we find the BB+2PL model unphysical and exclude it in the following sections since its second power-law component with $\Gamma\sim0.4$ exceeds the spin-down power of Geminga. 

We find any additional continuum component is either not statistically required or does not yield reasonable results. 
For example, although an additional (third) blackbody component to the 2BB+PL model further improves the fit, 
the model (3BB+PL) overpredicts the UV data by a factor of $\sim3$ and the fit radius is too large for reasonable NS models ($R=27\pm7$ km). Therefore we rule out the 3BB+PL model.  

\citet{kargaltsev_2005} pointed out that the \xmm\ spectrum hardens around $E\sim5$ keV. With the inclusion of \nustar\ data up to 20 keV we find that this hardening is even more pronounced. 
The hard power-law index of $\sim1.5$ above 3 keV reported in Section \ref{spec_3keV} supports this result. 

\begin{deluxetable*}{lcccccc}
 \tablecaption{Joint phase-averaged spectral fitting with \nustar\ and \xmm\ data }
 \tablehead{\colhead{Parameters} & \colhead{PL ($E>3$ keV)} & \colhead{BB+PL} & \colhead{2BB+PL} & \colhead{BB+2PL} & \colhead{BB+BKPL} } 
 \startdata
    $N_{\rm H}$ [$10^{20}$~cm$^{-2}$] & --- & $1.31\pm0.21$ & $1.54\pm0.26$ & $2.16\pm0.28$ & $1.93\pm0.24$  \\
    $kT_1$ [eV]\tablenotemark{a} & --- & $44.4\pm0.6$  & $44.0\pm0.8$ &  $41.6\pm0.8$ &  $42.4\pm0.6$  \\
    $R_1$ [km]\tablenotemark{b} & --- & $10.7\pm0.8$ & $11.4\pm1.1$ & $15.1\pm1.6$ & $13.7\pm1.2$  \\
    $kT_2$ [eV]\tablenotemark{a} & --- & --- & $195\pm14$ & --- & ---  \\
    $R_2$ [m]\tablenotemark{b} & --- & --- & $45\pm7$  &  --- & --- \\
    $\Gamma_1$ & $1.44\pm0.06$  & $1.90\pm0.02$ & $1.70\pm0.04$ & $2.15\pm0.08$ & $2.04\pm0.03$ \\
    $N_{\rm PL1}$\tablenotemark{c}   & $7.7\pm0.1$ & $6.0\pm0.3$  & $7.8\pm0.3$ & $8.0\pm0.1$ & $3.7\pm0.4$ \\
    $\Gamma_2$ & --- & ---  & ---  & $0.37\pm0.44$ & $1.42\pm0.07$ \\
    $N_{\rm PL2}$, $E_{\rm break}$\tablenotemark{d} & --- &  --- &  --- & $0.24\pm0.24$ & $3.4\pm0.3$ \\
    $\chi^2/dof$ & 0.968 & 1.063 & 0.991 & 0.889 & 0.904 \\
    dof & 121 & 449  & 447  &  446 &  446 
\enddata
 \tablenotetext{a}{$kT_1$ and $kT_2$ are the best-fit temperatures for the 1st and 2nd blackbody components, respectively.} 
 \tablenotetext{b}{$R_1$ and $R_2$ are the best-fit radii for the 1st and 2nd blackbody components, respectively. A distance of 250 pc from \citet{faherty_2007} is assumed. The uncertainty on the measured distance (250$^{+120}_{-62}$ pc) is not taken into account. }
 \tablenotetext{c}{The units of $10^{-5}$ ph~cm$^{-2}$~s$^{-1}$~keV$^{-1}$ at $E=1$ keV. }
  \tablenotetext{d}{Break energy ($E_{\rm break}$ [keV]) for BB+BKPL model. Power-law flux normalization ($N_{\rm PL2}$ [$10^{-5}$ ph~cm$^{-2}$~s$^{-1}$~keV$^{-1}$] at $E=1$ keV) for the other models.}
 \label{index_table}
\end{deluxetable*}

\section{Pulse profile and phase-resolved spectral analysis}
\label{phase_spec}

\begin{figure}[b]
\centerline{
\hfill
\includegraphics[angle=270,width=1.0\linewidth,clip=]{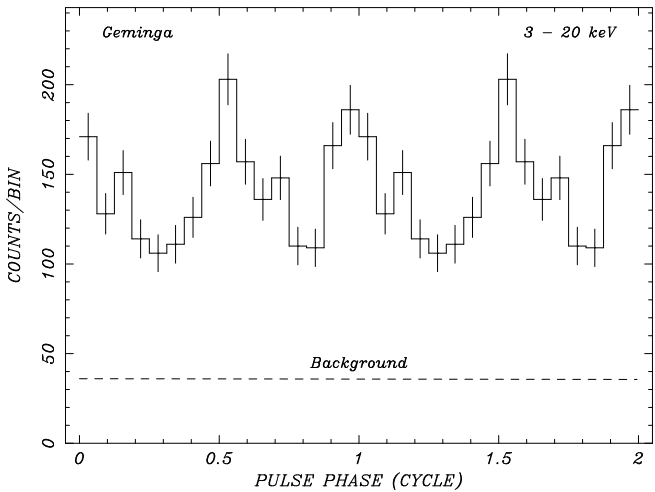}
\hfill
}
\caption{\textit{NuSTAR} pulse profile of the Geminga pulsar in the $3-20$ keV band. Photon arrival times
were folded on the ephemeris given in Table~\ref{tab:ephemeris}. Two cycles are displayed for clarity. }
\label{fig:nustarfolds}
\end{figure}

In this section, we present \nustar\ pulse profiles and phase-resolved spectral analysis. Using \nustar\ data of the pulsar B1509$-$58 as well as long-term \xmm\ and \fermi\ data of Geminga, 
we were able to determine the \nustar's absolute timing accuracy to better than 3~ms and derive a new ephemeris solution of Geminga (see the Appendix for details). 
We extracted source photons from an $44^{\prime\prime}$ radius aperture and
3-20 keV energy band, optimal for detecting pulsations.
Figure \ref{fig:nustarfolds} presents the \textit{NuSTAR} pulse
profile of Geminga folded on the \textit{Fermi} ephemeris (Table~\ref{tab:ephemeris} in the Appendix).
The intrinsic pulsed fraction in the 3-20~keV
\textit{NuSTAR} detection band is $f_p \approx 43\%$
(Figure~\ref{fig:nustarfolds}).
The broadband profile contains $1,850$
background subtracted counts collected during $154$~ks of livetime
that spanned the $8$-day data set.  The signal strength of
$Z^2_3=82.3$ has a negligible false detection probability ($\wp =                                                                                                       
1.2\times10^{-13}$).

\begin{figure}[b]
\centerline{
\hfill
\includegraphics[height=1.0\linewidth,angle=270, clip=]{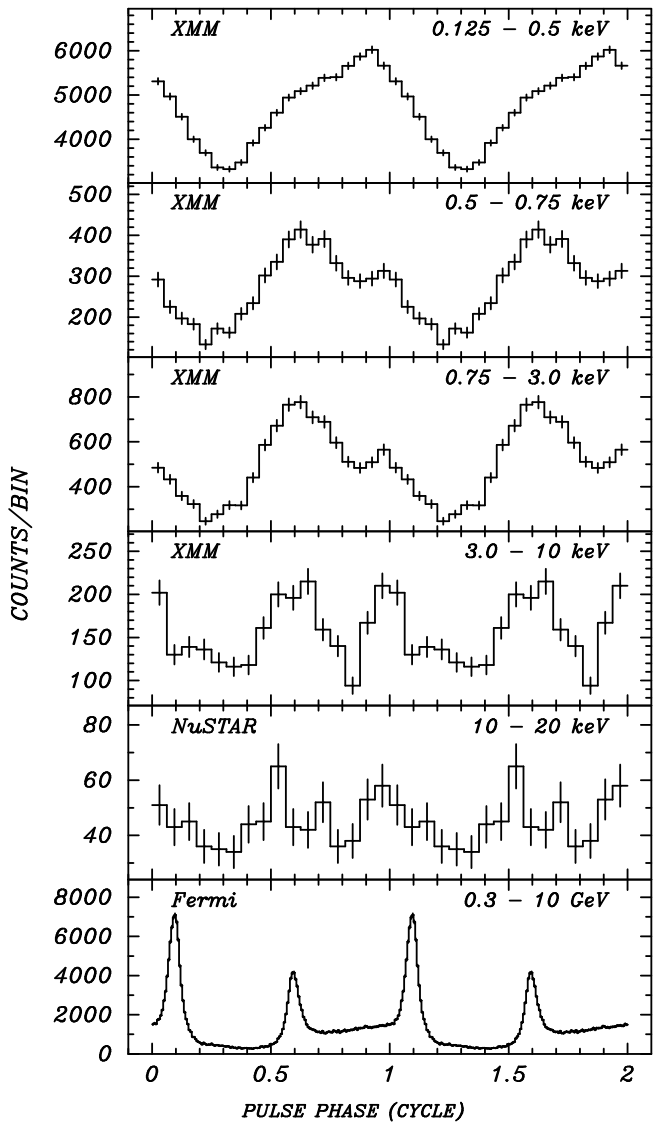}
\hfill
}
\caption{Pulse profiles of the Geminga pulsar from several missions
  and epochs, \xmm: 0.125-0.5~keV, 0.5-0.75~keV, 0.75-3.0~keV,
  3.0-10.0~keV, \nustar: $10-20$~keV and \fermi:
  0.3-10~GeV from top to bottom, respectively. Photon arrival times
  for \nustar\ and \fermi\ data sets are folded on the \fermi\
  ephemeris given in Table~\ref{tab:ephemeris}.  \fermi\ profile
  includes data that span from the end of the \xmm\ epoch (2008 Aug
  4) to the \nustar\ (2012 Oct 11) observations. Photon arrival times
  for \xmm\ data sets are folded on the \xmm\ ephemeris of 
  Table~\ref{tab:ephemeris}. Two cycles are displayed for clarity.}
\label{fig:pulseprofile}
\end{figure}

For comparison, Figure \ref{fig:pulseprofile} presents \xmm\ pulse
profiles in four energy bands, along with the \nustar\ (10-20 keV) and
\fermi\ (0.3-10 GeV) pulse profiles, all folded on the \textit{Fermi}
ephemeris given in Table~\ref{tab:ephemeris}.  We verified that the
two broad peaks in the \nustar\ pulse profile are statistically
identical to the \xmm\ profile in the overlapping 3-10~keV energy
band.

Given that the double-peaked pulse profile is persistent over the 3 to 20 keV band where non-thermal emission is dominant, it is 
possible that the observed spectral hardening is due to phase variation of the 
non-thermal component. Indeed, \citet{jackson_2005} demonstrated that the photon index of the non-thermal component varies strongly with phase.   
First, we performed phase-resolved spectroscopy using only the \xmm\ EPIC-PN data. 
We extracted \xmm\ EPIC-PN spectra from 10 phase intervals of $\Delta\phi = 0.1$ width using the ephemeris in Table~\ref{tab:ephemeris} (see Figure \ref{fig:pulseprofile} 
for the folded lightcurves). We followed the phase-integrated 
spectral analysis regarding background subtraction, spectral binning and statistical tests.   
Following \citet{jackson_2005}, we fit a BB+PL model to each of the phase-resolved spectra in 0.25-10 keV band with the absorption column $N_{\rm H}$ fixed to the value from the phase-integrated 
spectral analysis 
($N_{\rm H} = 1.3\times10^{20}$~cm$^{-2}$, see Table \ref{index_table}). Spectral fitting  with the BB+PL model was adequate for all phase-resolved spectra. 

\begin{figure*}
\centerline{
\hfill
\psfig{figure=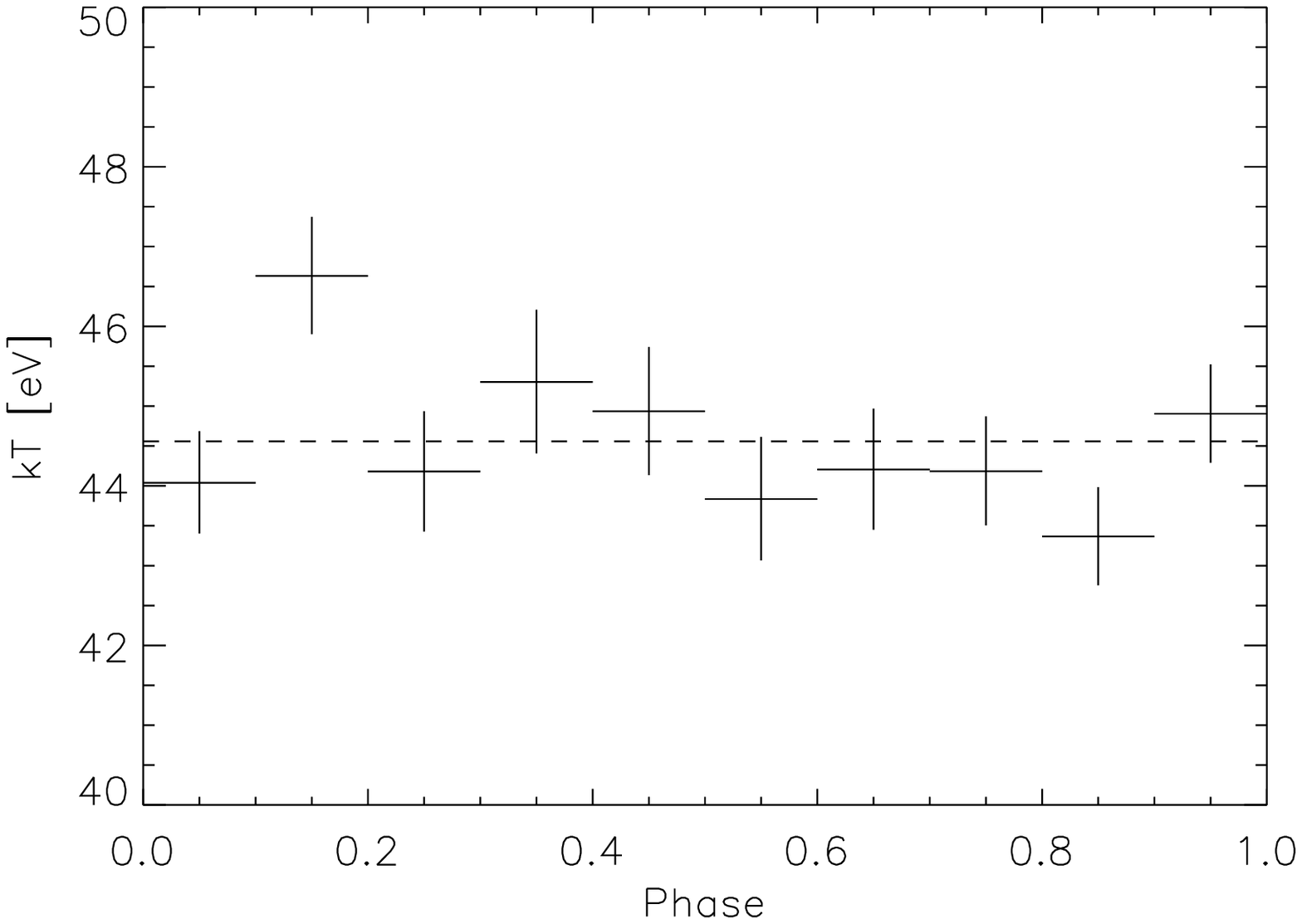,height=0.35\linewidth}
\hfill
\psfig{figure=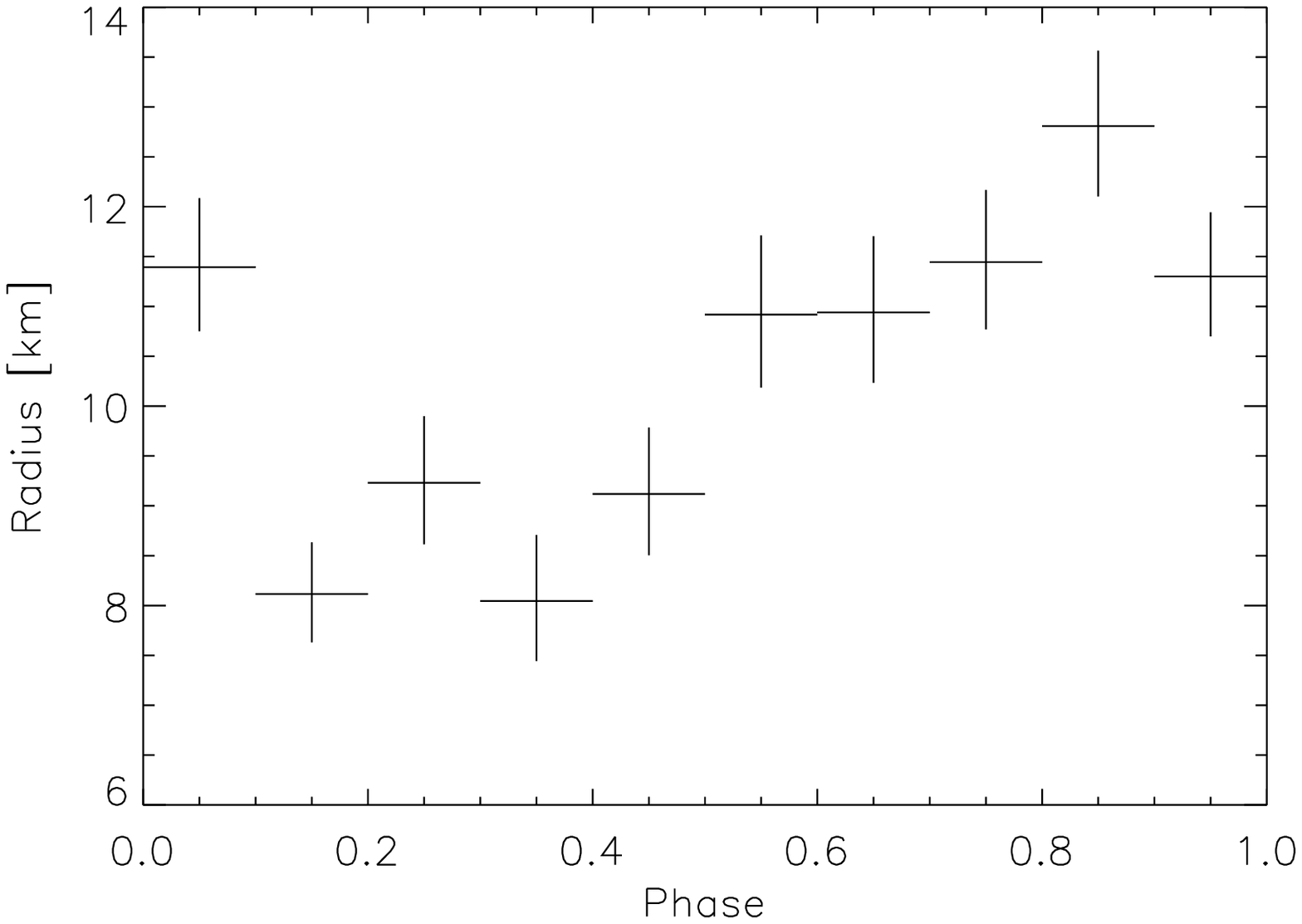,height=0.35\linewidth}
\hfill
}
\caption{Phase variation of the fitted blackbody temperature (left) and radius at $d = 250$~pc (right). \xmm\ data in the 0.25-10 keV band were used for spectral fitting. 
The phase zero is identical to that of the folded lightcurves in Figure \ref{fig:pulseprofile}. }
\label{fig:phase_vary1}
\end{figure*}

\begin{figure*}
\centerline{
\hfill
\psfig{figure=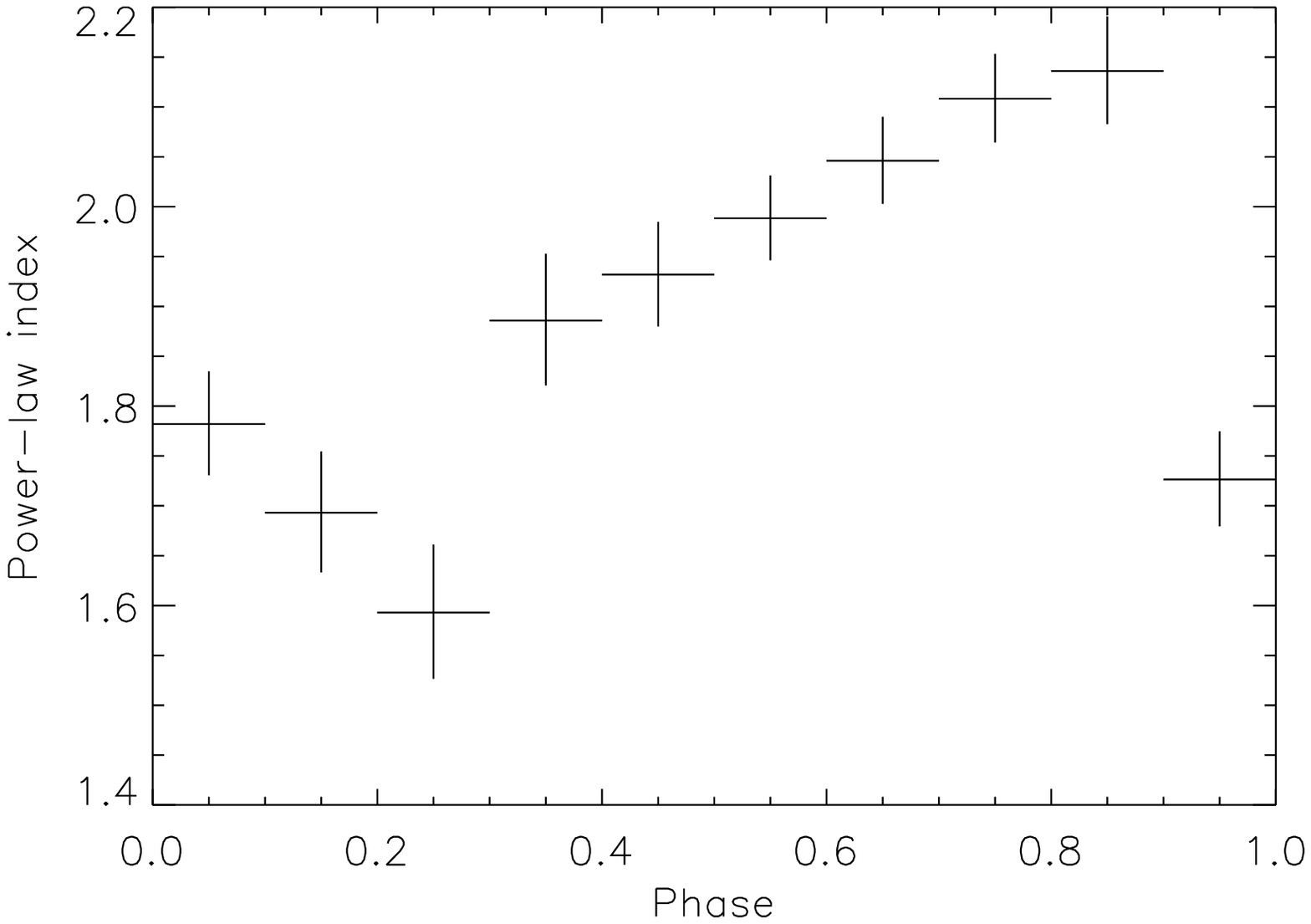,height=0.35\linewidth}
\hfill
\psfig{figure=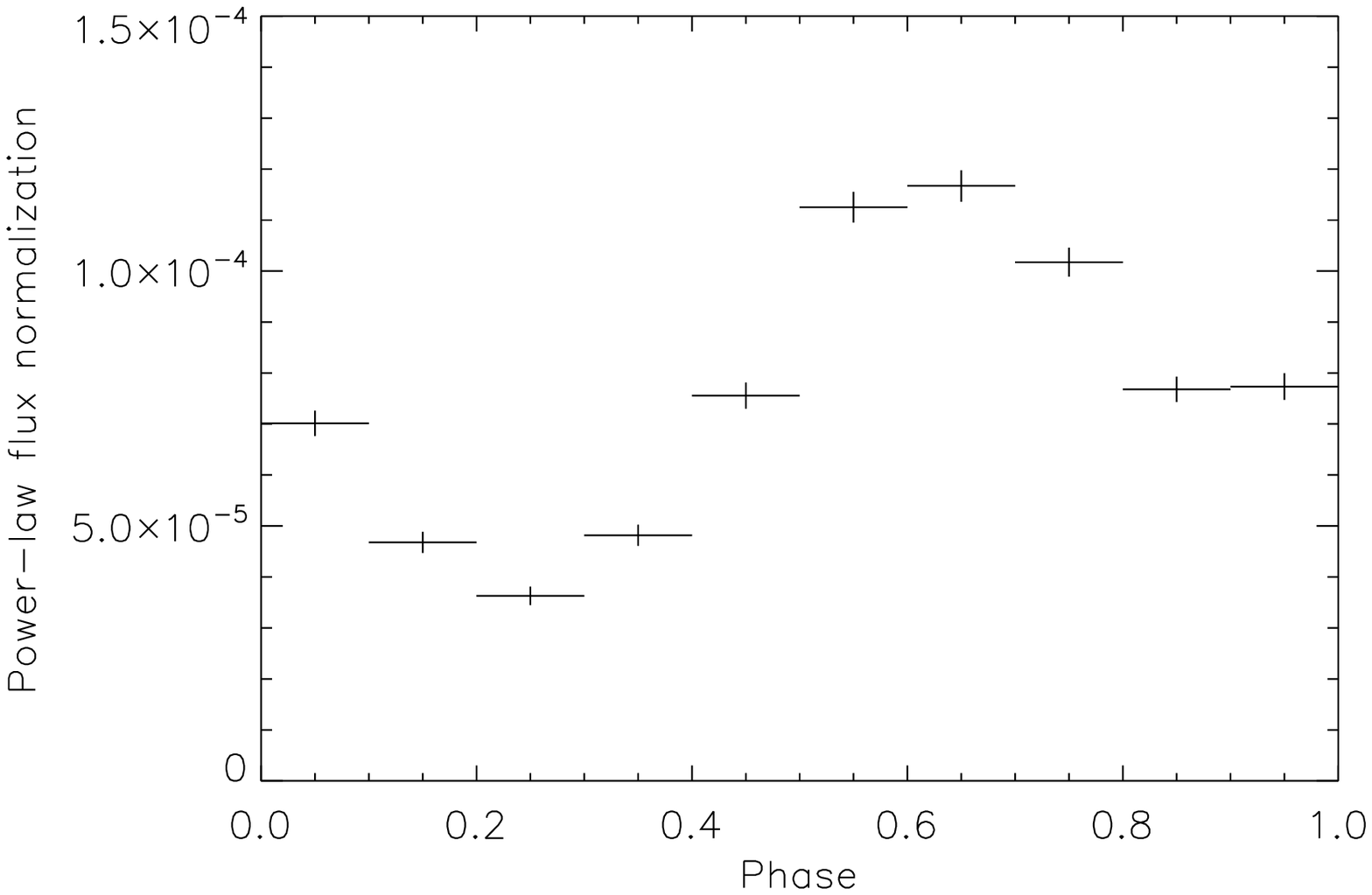,height=0.35\linewidth}
\hfill
}
\caption{Phase variation of the best-fit power-law index (left) and flux normalization [ph~cm$^{-2}$~s$^{-1}$~keV$^{-1}$] at $E=1$ keV (right) in the 0.25-10 keV band. 
\xmm\ data in the 0.25-10 keV band were used for spectral fitting. The phase zero is identical to that of the folded lightcurves in Figure \ref{fig:pulseprofile}.}
\label{fig:phase_vary2}
\end{figure*}

We then let $kT$, $\Gamma$ and the flux normalizations vary freely.  
The blackbody temperature remains nearly constant around $kT=44$~eV (left panel in Figure \ref{fig:phase_vary1}).  
 On the other hand, the power-law index strongly varies with phase (Figure \ref{fig:phase_vary2}) from $\Gamma = 1.59\pm0.06$ at $\phi = 0.2-0.3$ to $\Gamma=2.14\pm0.06$ at $\phi = 0.8-0.9$, 
confirming the results of \citet{jackson_2005}.
It is noteworthy that the non-thermal spectrum is hardest when the flux normalization is at its minimum ($\phi=0.2-0.3$).

We also studied the phase variation of the \nustar\ spectra.  
Given the limited statistics of the \nustar\ data, we extracted two phase-resolved \nustar\ spectra from $\phi = 0.0-0.3$ and $0.9-1.0$ (phase A) 
and $\phi = 0.3-0.9$ (phase B). Phase A and B represent the phase intervals where the \xmm\ power-law indices are harder ($\Gamma=1.5-1.8$) and softer ($\Gamma=1.9-2.2$),  
respectively. 
Since the non-thermal component completely dominates above 3 keV (even for the 2BB+PL model), we fit a single PL model to the 
\nustar\ spectra in three different energy bands. In the 3-20 keV band, both the phase A and B \nustar\ spectra fit to $\Gamma = 1.4 \pm0.1$. In the 3-10 keV band, phase A 
and B \nustar\ spectra yield $\Gamma = 1.4\pm0.2$ and $1.8\pm0.2$, similar to the phase variation found in the \xmm\ data. On the other hand, in the 5-20 keV band, phase A and 
B \nustar\ spectra exhibited harder power-law indices with $\Gamma = 1.4\pm0.2$ and $1.2\pm0.2$, respectively. 
This suggests that there is another non-thermal component emerging above 5 keV where we observed spectral hardening in the phase-integrated spectral analysis.


\section{Discussion}
\label{discussion}

\subsection{Thermal emission}

Conventionally, X-ray spectra of middle-aged pulsars have been interpreted as a combination of thermal and non-thermal emission. Thermal emission is thought to have two components --- a cold temperature component from the NS
surface and a hot temperature component from the heated polar caps. Two middle-aged pulsars, PSR B0656+14 and PSR B1055$-$52, exhibit two thermal components and a non-thermal component \citep{deluca_2005}. 
It has also been a common exercise to fit the X-ray spectra of the Geminga pulsar to either one or
two blackbody components \citep{halpern_1993, halpern_1997, jackson_2002, caraveo_2004, deluca_2005, jackson_2005, kargaltsev_2005}. 

Our spectral analysis rules out the hydrogen atmosphere model since it overpredicts the UV flux and the fit radius is too large for NS. Similarly, a model with three 
blackbody components is ruled out. Our best-fit models have either a single blackbody 
component (requiring a break in the non-thermal component) or two blackbody components (2BB+PL model).  
Our spectral analysis showed a second blackbody component is not present, or is faint, with a bolometric luminosity of 
 $L_x \sim 9\times10^{29}$ erg~s$^{-1}$ --- this is smaller than those of PSR B0656+14 and PSR B1055$-$52 by more than an order of magnitude.  

Based on previous theoretical work, hot polar caps can be faint either because they may not be fully visible to an observer or they are not sufficiently heated to 
emit observable thermal X-rays.   
\citet{cheng_1999} studied various thermal components including hot polar caps heated by returning current from the outer gaps and found that the 
visibility of these thermal components and their relative strengths compared to non-thermal emission strongly depends on the viewing angle 
 between the dipole magnetic axis and the observer's line of sight    
and inclination angle between the rotational and the dipole magnetic axes. Indeed, \citet{cheng_1999} predicted that hard thermal X-rays from hot polar caps might not be 
visible because the viewing angle is larger than the inclination angle. This was also independently concluded 
by \citet{romani_1995} who applied their gamma-ray emission models to the EGRET 
spectra and lightcurves of the Geminga pulsar. 

Alternatively, \citet{wang_1998} argued that relativistic electrons and positrons traveling from the outer gaps toward polar caps may undergo very efficient cooling by resonant inverse Compton scattering on 
radio photons, in addition to cooling by curvature radiation. However, in this scenario, it is not clear why Geminga is unique with fainter hot polar cap emission compared to other middle-aged pulsars.


\subsection{Non-thermal spectral hardening and broadband spectral energy distribution}

Most pulsar emission models attribute non-thermal X-ray emission to electron synchrotron radiation in the magnetosphere. \citet{wang_1998} predicted a power-law spectrum with $\Gamma\sim1.5$ in the 
X-ray band.  
Our spectral analysis suggests a spectral hardening at $E\sim5$ keV in the non-thermal component. 
Interestingly, the Vela pulsar, another middle-aged pulsar with a spin-down age of $10^4$ years, exhibits similar spectral hardening in the pulsed spectra obtained by \textit{RXTE} 
\citep{harding_2002}. 

\begin{figure*}[t]                
\includegraphics[width=0.75\linewidth]{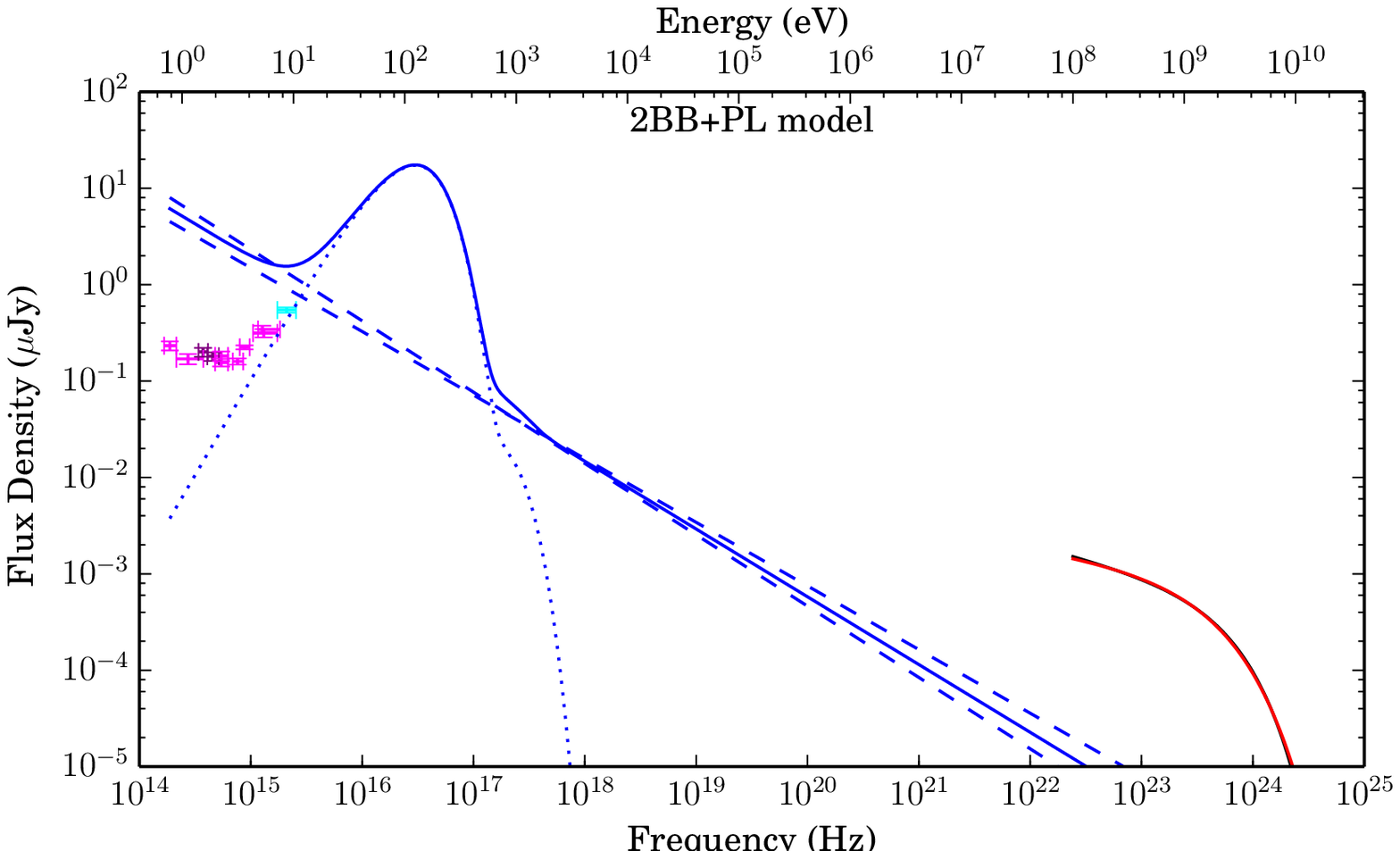}

\includegraphics[width=0.75\linewidth]{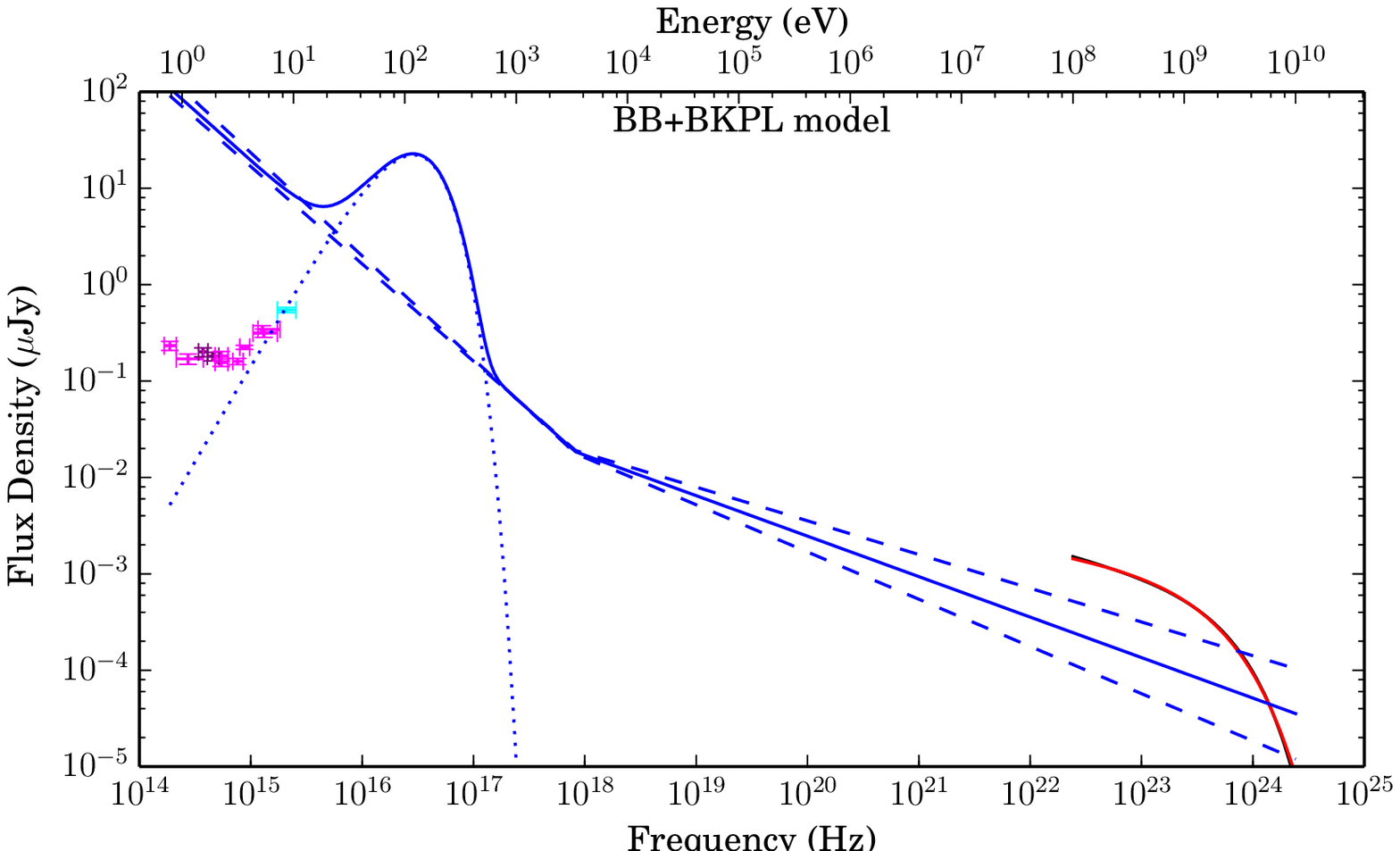}
\figcaption{Spectral energy distribution of the Geminga pulsar for two different model fits (top: 2BB+PL, bottom: BB+BKPL). 
Magenta points are {\it HST} imaging fluxes, purple points show Subaru imaging fluxes, and the cyan point is from a power-law fit to HST/STIS data, 
all of which have been dereddened \citep{kargaltsev_2005, shibanov_2006}. 
The red lines are an exponentially cutoff power-law fit to the \fermi\ phase-integrated spectrum \citep{abdo_2010}. 
The blue lines are the best fit models to \nustar\ and \xmm\ data between 0.3 and 20 keV. 
The dotted lines indicate thermal components, while the dashed lines indicate 68\% error envelopes of the non-thermal components.}
\label{fig:SED}
\end{figure*}

After re-analyzing archival data from {\it Spitzer Space Telescope}, \citet{danilenko_2011} studied the non-thermal emission of the Geminga and Vela pulsars across the mid-infrared (MIR), optical and X-ray bands.
For both pulsars as well as the Crab pulsar, non-thermal spectra in the optical band are significantly flatter than those in the X-ray band \citep{danilenko_2011}.
In all our best-fit models, an extension of the non-thermal spectra from the X-ray to lower frequencies over-predicts the optical fluxes (Figure 9). 
Therefore, it is evident that some spectral flattening takes place somewhere between $\sim0.05-0.5$ keV.

This spectral evolution between the optical and X-ray band is conceivable from a theoretical point of view.
\citet{wang_1998} pointed out that the electron cyclotron cutoff energy may be $E_c\sim0.1$ keV in the magnetosphere where the magnetic field strength ($B\sim10^{10}$ G) is significantly weaker than on the
NS surface ($B_S=1.6\times10^{12}$ G). Below the cyclotron cutoff energy, which was estimated to be $\sim0.02-0.5$ keV, the spectrum should have the canonical low-energy synchrotron index of $\Gamma \sim 
2/3$, much harder than the X-ray synchrotron spectra with $\Gamma \sim 1.5$ \citep{wang_1998}.

Towards higher energies, an extension of the BB+BKPL model ($\Gamma_2\sim1.4$) is roughly consistent with the phase-averaged \textit{Fermi} spectrum with $\Gamma\sim1.3$ \citep{abdo_2010}, while the 
2BB+PL is inconsistent with direct extrapolation to the \fermi\ band (see Figure 9). 
However, most theoretical models predict X-ray synchrotron spectra will become weaker in the MeV range
until different emission mechanisms (e.g. curvature radiation and inverse Compton) emerge toward the GeV band 
\citep{romani_1996}; this is indeed what was observed in the pulsed spectra of the Crab and Vela pulsars. \citet{wang_1998} estimated a high-energy cut-off of X-ray synchrotron spectra at $\sim5$ MeV. 
Some models also predict that one could observe synchrotron, inverse Compton, and curvature radiation dominating at different energies, and therefore expect multiple power-law components between the X-ray and GeV 
bands \citep{harding_2008}. None of these models argues that the slope of the non-thermal X-rays (which is due to synchrotron radiation) and that of the GeV 
gamma-rays should be the same. 

In summary, our spectral analysis confirms the spectral hardening at $E\sim5$ keV, and indicates that a comparison between the optical and X-ray non-thermal spectra requires a spectral flattening toward low energy between 
$\sim0.05$ and 0.5 keV. Thus, the Geminga pulsar should have two spectral breaks in its multi-wavelength non-thermal spectrum, in addition to spectral evolution from the X-ray to GeV bands as predicted by
a handful of pulsar 
emission models. The multiple spectral break scenario argues against the view of \citet{durant_2011} where a single power-law model might account for the multi-wavelength non-thermal spectra of middle-aged pulsars.


\section{Conclusion}
\label{conclusion}

Our 150 ks \nustar\ observation of the Geminga pulsar detects pulsed emission above 10 keV for the first time. The power-law spectrum and the double-peaked pulse 
profile, previously seen in the 3--10 keV soft band, persist above 10 keV. 
By combining \nustar\ and archival \xmm\ data, our broadband spectroscopy from 0.2 to 20 keV 
is able to constrain both the thermal and non-thermal emission from the pulsar. Our broadband spectral analysis from NIR to the hard X-ray band detects 
spectral hardening at $E\sim5$ keV and also indicates spectral flattening between the optical and hard X-ray bands, similar to what is seen in 
the Vela pulsar.  
It will be intriguing to observe other middle-aged pulsars with \nustar\ to search for spectral breaks in the (hard) X-ray band.

\acknowledgements

This work was supported under NASA Contract No. NNG08FD60C, and made use of data from the \nustar\ mission, a project led by the California Institute of Technology,
managed by the Jet Propulsion Laboratory, and funded by the National Aeronautics and Space Administration. We thank the \nustar\ Operations, Software
and Calibration teams for support with the execution
and analysis of these observations. This research has
made use of the \nustar\ Data Analysis Software (NuSTARDAS) jointly developed by the ASI Science Data Center (ASDC, Italy) and the California Institute of
Technology (USA). E.V.G. acknowledges support from NASA/Fermi grant NNX12AO89G and NASA/Chandra grant G03-14066X. V.M.K. acknowledges support from an NSERC Discovery Grant, the FQRNT Centre de Recherche Astrophysique du
Qu´ebec, an R. HowardWebster Foundation Fellowship from the
Canadian Institute for Advanced Research (CIFAR), the Canada
Research Chairs Program, and the Lorne Trottier Chair in Astrophysics
and Cosmology. A.M.B. acknowledges the support
by NASA grants NNX10AI72G and NNX13AI34G.

\appendix

\section{A. Timing analysis and derivation of a new ephemeris of Geminga}
\label{timing}

To allow a joint phase-resolved spectral analysis of the \xmm\ and the
\nustar\ Geminga data we require phase-connected timing solutions
spanning the two missions.  The \xmm\ data itself is suitable for
self-generating a phase-connected timing solution as the data were
obtained for this purpose. The ephemeris presented
here bridges the gap between the end of the EGRET
and the start of the \fermi\ mission.  As the \xmm\ observations do
not overlap with the \nustar\ data, and no published ephemeris is available
for that epoch, we generated a \fermi\ ephemeris that covers the end
of the \xmm\ observations to the \nustar\ data. In the following,
photon arrival times from all Geminga data sets were corrected to the
Solar System barycenter using the JPL D200 ephemeris calculated using
the {\it HST} coordinates of \citet{caraveo_1998} and the proper motion of
\citet{faherty_2007}, updated from \citet{caraveo_1996}, reproduced in Table~\ref{tab:ephemeris}.

Table~\ref{observations_table} lists all archival \xmm\ observations
for Geminga acquired in high time-resolution {\tt SmallWindow} mode.
Observational details for these data sets are presented in
Section~\ref{XMM_obs}. For our timing analysis we selected 0.2-10~keV source
photons from a $30\arcsec$ radius aperture centered on Geminga.  Extracted
data were initially folded at the period for the peak signal using the
$Z^2_1$ statistic and cross-correlated with an iterated high statistic
pulse profile to generate times-of-arrival (TOAs) for each data set.
These TOAs were fitted to a phase model with two frequency derivatives
initiated using the overlapping EGRET ephemeris.  This process was
iterated to produce the \xmm\ ephemeris presented in
Table~\ref{tab:ephemeris}.

To overlap with the \xmm\ ephemeris, we analyzed \fermi\ data covering
the mission start to the \textit{NuSTAR} epoch.  Data were obtained
from the \textit{Fermi/LAT} archive and photons selected from the 200
MeV to 10 GeV range within a $1.3^{\circ}$ aperture centered on the
pulsar. These photons were filtered to include only events tagged as
{\tt Pass~7} ``Source'' photons and restricted to a maximum zenith
angle of $\phi<100^{\circ}$.  \fermi\ photon arrival times were binned
into $20$-day intervals and folded on the \xmm\ ephemeris to generate
times-of-arrival (TOAs) as described above.  These TOAs were fitted to
produce an iterative \fermi\ ephemeris presented in  Table \ref{tab:ephemeris}.

\begin{deluxetable*}{lc}
\tablecaption{Geminga Ephemeris}
\tablehead{
\colhead{Parameter} & \colhead{Value}
}
\startdata
Epoch of  Coordinates (MJD)\tablenotemark{a}  & 49794.0\\
R.A. (J2000)                                  & $06^{\rm h}33^{\rm m}54^{\rm s}\!.153$ \\
Decl. (J2000)                                 & $ +17^{\circ}46^{\prime}12^{\prime\prime}\!.91$\\
R.A. proper motion, $\mu_{\alpha}\,{\rm cos}\,\delta$  & $142.2\pm 1.2$ mas yr$^{-1}$ \\
Decl. proper motion, $\mu_{\delta}$             & $107.4\pm 1.2$ mas yr$^{-1}$ \\
\cutinhead{\xmm\ Timing Solution (2002 Apr 5 - 2009 Mar 10)}
Epoch of ephemeris (MJD)\tablenotemark{b} & 53630.0 (TDB)\\
Span of ephemeris (MJD)                       & 52,369-54,900\\
Frequency, $f$                                & $4.21758680107(16)$~Hz \\
Frequency derivative, $\dot f$                & $-1.952196(17) \times 10^{-13}$ Hz s$^{-1}$ \\
Frequency second deriv., $\ddot f$            & $-3.20(90) \times 10^{-25}$ Hz s$^{-2}$ \\
Period, $P$                                   & $0.2371024112050(90)$~s \\
Period derivative, $\dot P$                   & $1.0974768(96) \times 10^{-14}$\\
Period second deriv., $\ddot P$               & $1.90(51) \times 10^{-26}$ \\
\cutinhead{\fermi\ Timing Solution (2008 Aug 04 - 2012 Oct 12)}
Epoch of ephemeris (MJD)\tablenotemark{b}     & 55440.0 (TDB)\\
Span of ephemeris (MJD)                       & 54,682 - 56,212\\
Frequency, $f$                                & $4.217556269653(16)$~Hz \\
Frequency derivative, $\dot f$                & $-1.9521993(37) \times 10^{-13}$~Hz~s$^{-1}$ \\
Frequency second deriv., $\ddot f$            & $5.64(33) \times 10^{-25}$~Hz~s$^{-2}$ \\
Period, $P$                                   & $0.23710412761897(90)$~s \\
Period derivative, $\dot P$                   & $1.0974946(21) \times 10^{-14}$\\
Period second deriv., $\ddot P$               & $-3.07(19) \times 10^{-26}$\\
\enddata
\tablenotetext{a}{Coordinates used to barycenter photon arrival times,
  based on HST measurements \citep{caraveo_1998} and the proper motion
  results of \cite{faherty_2007}.}  \tablenotetext{b}{Phase zero in
  Figure~\ref{fig:nustarfolds} and \ref{fig:pulseprofile}.}
\label{tab:ephemeris}
\end{deluxetable*}

\subsection{B. Preliminary Limits on the Absolute Timing Accuracy with \nustar}

The \nustar\ photon arrival times are corrected for spacecraft clock
drift with a typical RMS residual uncertainty of $\sim 2$~ms,
dominated by orbital temperature variations of
the clock. A high resolution \nustar\ absolute timing calibration is
underway using Crab observations to attempt to reduce these variations. For the present study, we verify that the
\nustar\ timestamps are of sufficient accuracy, in absolute Barycentric Dynamical Time (TDB) time,
to co-add phase-resolved \nustar\ Geminga spectra with those obtained
with the \xmm\ mission. For this purpose we have analyzed near
simultaneous \swift\ and \nustar\ observations of the $151$~ms pulsar
B1509$-$58 in SNR MSH~15$-$52. 
Any time offset is already known to be less than $151$~ms from comparisons of \swift\ and 
\nustar\ observations of the 3.79~s pulsar SRG~J1745$-$2900 \citep{mori_2013,kaspi_2014}.

A \nustar\ observation of PSR~B1509$-$58 (ObsID 40024004002) was
obtained on 2013 June 7 to study the pulsar wind nebula MSH~15$-$5{\sl 2}.
The default pipeline processing resulted in a total of 45~ks of good
data. A short, 2.8~ks \swift\ XRT observation (ObsID 00080517001) was
acquired 105~s prior to the end of the \nustar\ observation. The
default \swift\ pipeline was run on the two consecutive orbits of data
(7,340~s span) taken in window timing (WT) mode. This mode provides
$1.78$~ms timestamps with a clock-drift corrected MJD absolute TDB
\swift\ times better than $0.2$~ms \citep{cusumano_2012}.  For this
study it is not necessary to apply the UVOT attitude correction to try
and improve the \swift\ absolute time accuracy. Both data sets were
barycentered using the JPL DE200 Solar System ephemeris and the
\chandra\ (ObsID 3833) determined coordinates (J2000) $15^{\rm                                                                                                                                                                   
  h}13^{\rm m}55^{\rm s}\!.66$, $                                                                                                                                                                                                
-59^{\circ}08^{\prime}09^{\prime\prime}\!.2$ (epoch MJD 52930).

\begin{figure}[b]
\centerline{
\hfill
\includegraphics[angle=270,width=0.5\linewidth,clip=]{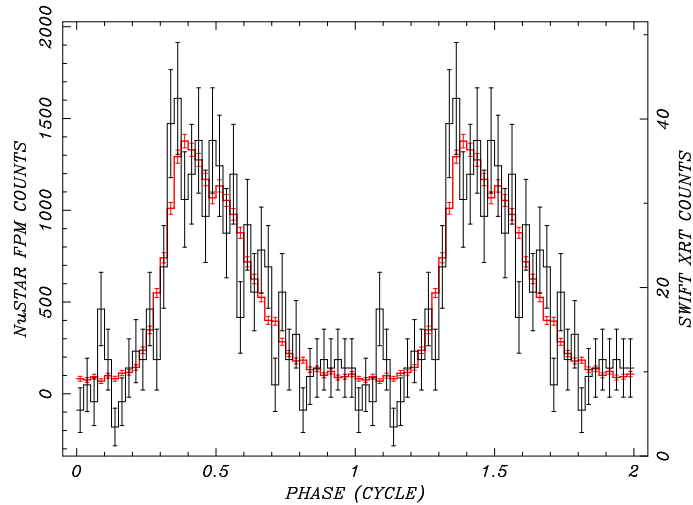}
\hfill
}
\caption{\nustar\ (red line) and \swift\ (black line) pulse
  profile of PSR~B1509$-$58 in the overlapping 3-10~keV energy band
  folded in 40 phase bins on the same ephemeris. The data were acquired
  nearly simultaneously in time. The calculated phase lag between the two profiles is less
  than one phase bin. Two cycles are displayed for clarity.}
\label{fig:b1509_lag}
\end{figure}

The \nustar\ and \swift\ observations yield a total of $22,629$ and
$796$ counts, respectively, extracted from an aperture of $r=0.5'$
radius in the overlapping 3-10~keV energy band.
Figure~\ref{fig:b1509_lag} compares the pulse profile from the two
missions folded on the same period and period derivative at epoch MJD 56450.885504116, in $40$ phase
bins.  The period was determined from the peak signal in the highly
significant \nustar\ data set using the $Z^2_5$ statistic and the
period derivative was obtained from the radio ephemeris reported in
\cite{martin_2012}.  Cross-correlating the two profiles yields a phase lag
corresponding to a relative time offset of $\Delta \phi =                                                                                                                                                                        
1.0\pm2.0$~ms, comparable with the residual uncertainty in the
\nustar\ clock drift correction.  This measurement represents an
upper-limit since the ($1\sigma$) error is dominated by the photon
counts of the short \swift\ exposure. We conclude that the present
data are consistent with the result of no measurable phase offset
compared to the calibrated \swift\ clock. This verifies that the
\nustar\ absolute time is sufficiently accurate to phase align co-added
\xmm\ and \nustar\ spectral data for Geminga in 10 phase bins.

\bibliographystyle{apj}
\bibliography{thebib}

\end{document}